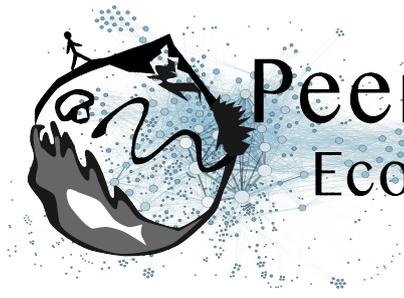

RESEARCH ARTICLE

# A reply to 'Ranging Behavior Drives Parasite Richness: A More Parsimonious Hypothesis'

Marie JE Charpentier, Peter M Kappeler

**A reply to "Ranging Behavior Drives Parasite Richness: A More Parsimonious Hypothesis"**

Charpentier MJE[1], Kappeler PM[2].

1. Institut des Sciences de l'Évolution de Montpellier (ISE-M) UMR5554, Univ. Montpellier, CNRS, IRD, EPHE, Montpellier, France

2. Behavioral Ecology and Sociobiology Unit, German Primate Center, Göttingen, Germany

# ABSTRACT

This preprint has been reviewed and recommended by Peer Community In Ecology (https://dx.doi.org/10.24072/pci.ecology.100001).

In a recent article, Bicca-Marques and Calegaro-Marques [2016] discussed the putative assumptions related to an interpretation we provided regarding an observed positive relationship between weekly averaged parasite richness of a group of mandrills (*Mandrillus sphinx*) and their daily path lengths (DPL), published earlier in the same journal (Brockmeyer et al., 2015). In our article, we proposed, *inter alia*, that 'the daily travels of mandrills could be seen as a way to escape contaminated habitats on a local scale'. In their article, Bicca-Marques and Calegaro-Marques [2016] proposed an alternative mechanism that they considered to be more parsimonious. In their view, increased DPL also increases exposure to novel parasites from the environment. In other words, while we proposed that elevated DPL may be a consequence of elevated parasite richness, they viewed it as a cause. We are happy to see that our study attracted so much interest that it evoked a public comment. We are also grateful to Bicca-Marques and Calegaro-Marques [2016] for pointing out an obvious alternative scenario that we failed to discuss and for laying out several key factors and assumptions that should be addressed by future studies examining the links between parasite risk and group ranging. We use this opportunity to advance this discourse by responding to some of the criticisms raised in their discussion of our article. In this reply, we briefly contextualize the main object of criticism. We then discuss the putative parsimony of the two competing scenarios.

Little is still known about how wild animals organize their ranging behavior in response to the risks emanating from environmentally-transmitted parasites. In 2015, we published new data on group composition and patterns of male migration in wild mandrills and complemented this description of social organization with data on ranging behavior and home range use (Brockmeyer et al. 2015). Among many other results and conclusions, we suggested that mandrills may accept additional ranging costs to avoid heavily parasitized areas. Bicca-Marques and Calegaro-Marques [2016] subsequently questioned this interpretation of one of several correlative relationships reported in the second part of our article. By not acknowledging the main focus of our paper and the preliminary nature of the analysis of the group's ranging behavior, which was yet clearly stated, Bicca-Marques and Calegaro-Marques [2016] have created, in our view, a heavily distorted point of departure for their article. In addition, Bicca-Marques and Calegaro-Marques [2016] failed to mention that we also proposed an alternative interpretation of the observed relationship based on an interaction between food availability and parasite load. Thus, we resent the impression that the incomplete and biased depiction of our article by Bicca-Marques and Calegaro-Marques [2016], which may have created confusion in readers unfamiliar with the original study.

Bicca-Marques and Calegaro-Marques [2016] proposed that our interpretation of the observed positive relationship between daily path length (DPL) and parasite richness at the group level was based upon four implicit assumptions (detailed below). We think that the way these assumptions have been discussed is incomplete and deserving of additional comments.

Assumption 1 relies on the supposedly non-pathogenic nature of the studied protozoan taxa and on the absence of any mention of signs of sickness in infected mandrills. The health and fitness effects of these protozoan taxa are largely under-studied (and remain completely unknown in wild mandrills). While most of them do not cause any evident signs of sickness, this does not *a priori* and necessarily equate with an absence of any fitness effects. Second, some of these protozoan taxa have been shown to impact health, especially in immunocompromised individuals (e.g., *Balantidium coli*: Schuster and Ramirez-Avila, 2008). *E. histolytica*, which represents 10% of infected individuals against 90% for *E. dispar* (Poirotte et al., 2017), is also clearly pathogenic (e.g., Stauffer and Ravdin, 2003).

We think that the key question is whether or not these parasites create strong enough selective pressures for avoidance mechanisms to have evolved. The absence of evident signs of sickness is not sufficient to conclude that this is not the case, i.e., the absence of evidence is not evidence of absence. For example, even at a low infestation level and without clear clinical symptoms, the growth rate of parasitized sheep is 50% inferior to that of dewormed individuals (Sykes and Coop, 1977), suggesting that energy is allocated to physiological and immunological anti-parasite responses. Moreover, several parasites or viruses, traditionally considered as benign, have been found to be highly virulent following long-term population monitoring. The SIV (Simian Immunodeficiency Virus) infecting chimpanzees and other primates provides a striking example. Indeed, fitness data collected for more than 10 years on a large wild population of chimpanzees revealed that SIV has a substantial negative impact on the health, reproduction and lifespan of infected individuals, challenging the previous notion that almost all natural SIV infections were non-pathogenic (Keele et al., 2009). In mandrills, there is no conspicuous health effect of these protozoan infections, although we regularly observe cases of diarrhea (MJEC pers. obs.). In addition, protozoan richness is correlated to a suite of avoidance behaviors, probably to limit social transmission (Poirotte et al., 2017). Thus, it is probably premature to assume *a priori* that these parasites do not induce selective pressures strong enough for behavioral avoidance mechanisms to evolve.

Assumption 2 states that symptoms are more severe in multi-infected individuals or that the probability of hosting a pathogenic species is higher in these animals. There is evidence for multiplicative and unexpected effects of multi-infections in animals and humans (Vaumourin et al., 2015). For example, co-infections by nematodes cause more severe pathologies in sheep than single infections (Steel et al., 1982; Sykes et al., 1988). Simultaneous infections with rotavirus and either *Giardia* or *Escherichia coli* result in a greater risk of having diarrhea in human populations than expected if the co-infecting pathogens act independently of each other (Bhavnani et al., 2012). Finally, and as stated by Bicca-Marques and Calegaro-Marques [2016], the second part of this assumption is simply a question of probability.

Assumption 3 is related to group movements and decision-making processes. Because the mechanisms underlying group movements have not been studied in this or any other mandrill group, this point is empirically open. Group travel is either based on shared-decision processes (which would support the interpretation offered by Brockmeyer et al., 2015; see also: Strandburg-Peshkin et al., 2015) or on decision-makers that are either infected at the average population level (which would make the same prediction) or able to perceive an increase in parasite richness at the group-level. Bicca-Marques and Calegaro-Marques [2016] stated that in the absence of evident signs of sickness, the latter part of the assumption is unlikely, but this claim ignores a large literature on perception of parasitism, for example, *via* olfactory cues (Prugnolle et al., 2009; Poirotte et al., 2017).

Finally, assumption 4 states that repeated use of a smaller area increases the risk of exposure to novel parasites or facilitates the spread of parasites between group-members. We contend that this assumption is not as strong as proposed and that the associated discussion is incomplete. First, Bicca-Marques and Calegaro-Marques [2016] proposed that the risk of encountering novel parasites is higher when ranging farther (their assumption), citing studies (e.g., Benavides et al., 2012; Han et al., 2015) that were not designed to distinguish between these two alternatives, however. Second, Bicca-Marques and Calegaro-Marques [2016] proposed that the low prevalence of three protozoa taxa in our study population is "compatible with the idea of foragers experiencing a higher likelihood of encountering novel parasite species when traveling longer distances". The way we analyzed our data, by considering concurrently weekly averaged parasite richness and weekly DPL, challenges this view, however. Indeed, there is a time gap between parasite exposure and successful establishment in the host (6-8 days for protozoa; Golvan 1983). While, the mechanism proposed by Bicca-Marques and Calegaro-Marques [2016] would predict a better fit with parasite richness estimated about a week before DPL measurements, our analyses were designed to test for the mechanism we proposed. In addition, the low prevalence of these parasites may not only reflect their probability of being encountered in the environment; there are other possible physiological explanations for their low prevalence related, for example, to their low transmissibility or the mandrills' susceptibility. Bicca-Marques and Calegaro-Marques [2016] further discussed the fact that the high prevalence of four other protozoa in the studied mandrills challenges the validity of the mechanism we proposed because high DPL would not reduce parasite spread between group-members. However, re-infection with common parasites increases individuals' burden and results in substantial costs in humans (e.g., Brooker et al., 2004) and other species (e.g., Ferguson et al., 2011). In other words, individuals should avoid being re-infected with common parasites, and escaping contaminated environments may allow just that.

Thus, the alternative explanation offered by Bicca-Marques and Calegaro-Marques [2016] was partly based on certain untested assumptions or ignorant of available evidence. While we do not question the fact that the alternative explanation offered by these authors could have been mentioned, we also think that our interpretation should not been dismissed at this stage of knowledge either. We agree that the mechanism proposed for the alternative explanation could be more parsimonious, but parsimony is not a necessary criterion for a complex trait to evolve. We hope that our comments will help to move this debate forward and that it will stimulate more empirical studies of the role of environmentally-transmitted parasites in shaping ranging and movement patterns of wild vertebrates.